\documentclass[a4paper,fleqn]{cas-sc}

\usepackage[utf8]{inputenc}
\usepackage[english]{babel}
\usepackage{graphicx}
\usepackage{bm}
\usepackage{xcolor}
\usepackage{tcolorbox}

\usepackage[authoryear]{natbib}

\def\tsc#1{\csdef{#1}{\textsc{\lowercase{#1}}\xspace}}
\tsc{WGM}
\tsc{QE}
\ExplSyntaxOn
\keys_set:nn { stm / mktitle } { nologo }
\ExplSyntaxOff

\begin{document}

\let\WriteBookmarks\relax
\def\floatpagepagefraction{1}
\def\textpagefraction{.001}

\shorttitle{}

\title [mode = title]{Dislocation distribution near a wall within the framework of the continuum theory of curved dislocations}

\author[1,2]{István Groma}[orcid=0000-0002-6644-1365]
\cortext[1]{Corresponding author}
\ead{groma@metal.elte.hu}
\credit{Conceptualization of this study, Theoretical foundation, Original draft preparation}
\affiliation[1]{organization={Institute for Solid State Physics and Optics, HUN-REN Wigner Research Centre for Physics},
            addressline={Konkoly T. út 29-33.},
            city={Budapest},
            postcode={1525},
            country={Hungary}}

\affiliation[2]{organization={Department of Materials Physics, E\"otv\"os Lor{\'a}nd University},
            addressline={P\'azm\'any P. stny. 1/A},
            city={Budapest},
            postcode={1117},
            country={Hungary}}

\author[2]{Dénes Berta}
\ead{berta.denes@ttk.elte.hu}
\credit{Numerical calculations, data analysis}

\author[2]{Lóránt Sándli}
\ead{sanbdli.lorant@ttk.elte.hu}
\credit{Numerical calculations, Data analysis}

\author[2,1]{Péter Dusán Ispánovity}
\ead{ispanovity.peter@ttk.elte.hu}
\credit{Conceptualization of this study, Data analysis}

\begin{abstract}
A recently proposed generalised continuum theory of curved dislocations describes the spatial and temporal evolution of statistically stored and geometrically necessary dislocation densities, as well as curvature. The dynamics follow from a scalar plastic potential that constrains the allowed velocity fields and leads to a phase‑field–like formulation with a nontrivial mobility function. Although conceptually related to strain‑gradient plasticity, the theory differs by introducing an intrinsic, evolving length scale given by the dislocation spacing.

In this paper, we determine three key material‑independent parameters of this continuum theory by quantitatively comparing its predictions with discrete dislocation dynamics (DDD) simulations. To achieve this, we impose a narrow impenetrable wall inside the simulation volume, which blocks dislocation motion and generates characteristic spatial variations of the dislocation density fields under external loading. We show that for this geometry, the continuum equations reduce to a form that can be solved efficiently via direct numerical integration. The resulting stationary distributions of total and geometrically necessary dislocation densities are then compared to extensive 2D and 3D DDD simulations. This comparison allows us to extract the parameters that govern the back‑stress, the density-gradient coupling, and the flow stress relation. Our results demonstrate that the continuum theory quantitatively captures the DDD‑observed structure of the dislocation pile‑up near the wall and therefore provides a reliable mesoscale description. The wall‑loading setup further serves as a benchmark problem to validate numerical implementations of the continuum theory in more general geometries.
\end{abstract}


\maketitle

\section{Introduction}
The plastic deformation of crystalline materials is controlled by the motion of a large number of dislocations. The typical dislocation density in deformed metals is in the order of  $\rho \sim 10^{14}$ m$^{-2}$, i.e. the average spacing between dislocation lines is about 100 nm. Therefore, even a deformed micron-sized sample contains a vast amount of strongly interacting dislocations. Thus, following the evolution of the dislocation network by discrete dislocation dynamics (DDD) simulations is computationally rather demanding
\citep{kubin1992,ghoniem1999,devincre2001,devincre2002,bulatov2006computer,arsenlis2007enabling,akhondzadeh2020dislocation,fan2021strain,berta2025identifying}.
Because of these limitations, for most problems, an appropriate continuum model should be suitable. However, due to the many degrees of freedom in the dislocation system, modelling the plastic deformation of crystalline materials in terms of dislocations requires handling the problem with statistical physics methods.  Nevertheless, existing traditional statistical physics methods cannot be directly applied because the dislocation motion is strongly dissipative and dislocations are flexible lines, which precludes their treatment as point-particles.

The development of a statistical continuum theory of dislocations was motivated by the discovery of the size-effects \citep{fleck1997advances} in the plastic response of samples with characteristic dimensions on the order of 10 $\mu$m or less. Despite several attempts to incorporate internal length scales into phenomenological continuum theories by adding strain-gradient terms to the stress \citep{zhu1997strain,aifantis1999,fleck,gurtin} there has been no satisfactory solution for general loading cases.

Another key feature observed is the formation of dislocation patterns during plastic deformation.  Since the early 1960s, several theoretical and numerical attempts have been suggested  based on analogies with other physical problems like spinodal decomposition \citep{holt1970}, internal energy minimisation \citep{hansen1986}, or chemical reaction-diffusion systems \citep{aifantis1985,pontes}. However, since they are not directly linked to the specific properties of individual dislocations, they are fundamentally phenomenological approaches.

Patterning also served as a significant  motivation for the development of DDD methods \citep{kubin1992,ghoniem1999,devincre2001,devincre2002}, but due to the long range dislocation-dislocation interactions, the simulations are computationally extremely expensive and poorly scalable. Thus, DDD is still limited to specific problems like irregular clusters or veins \citep{devincre2001,devincre2002,hussein2015}. Recently, El-Azab and coworkers \citep{xia2015computational,lin2020implementation} used a continuum formulation based on vector dislocation densities in large-scale numerical simulations, which can capture the evolution of dislocation patterns. However, this pseudo-continuum variant of DDD is a numerical rather than a fully theoretical model of dislocation patterning.

By a systematic coarse-graining of the evolution equation of individual dislocations, in a strongly simplified quasi two-dimensional system of straight parallel edge dislocations
a continuum theory was developed over the past 25 years \citep{groma1997link,zaiser2001statistical,groma2003spatial,groma2007dynamics,mesarovic2010,dogge2015,groma2016dislocation,valdenaire2016density} It has been successfully compared to DDD simulations \citep{groma2003spatial,yefimov2004comparison,groma2006debye,ispanovity2020emergence}. By now, it can be considered as a well-established theory for the 2D problem it addresses. Moreover, it has shown that the model can be formulated as a specific phase field theory \citep{groma2007dynamics,groma2010variational,groma2015scale,groma2016dislocation}. In contrast to many other phase field theories, the phase field functional in this case could be strictly derived from the statistical theory, and is not obtained on phenomenological grounds. The most important feature of the theory is that it predicts dislocation patterning, even though it was not ``designed'' for it \citep{groma2016dislocation,wu2018instability,ispanovity2020emergence}.

Given the fact that dislocations are moving curved flexible lines, an appropriate continuum theory should account for this.  The kinematic theory of the evolution of curved dislocations was developed by Hochrainer \emph{et al.}\citep{hochrainer2007three,sandfeld2010numerical,hochrainer2014continuum,hochrainer2015multipole}. The kinematics was initially derived in a 2+1D dimensional space, containing the line direction as an independent variable. A multipole expansion of the theory results in a formulation in terms of alignment tensors, which, in the case of only planar dislocations in parallel glide planes, is equivalent to a Fourier expansion \citep{groma2021dynamics}.

However, to obtain a closed theory, the velocity fields in the kinematic equations have to be given as a function of the dislocation densities. Based on the analogy with the 2D case \citep{groma2016dislocation} it is assumed that there is a scalar functional of the different fields, called ``plastic potential'', which cannot increase during the evolution of the system. This condition imposes a strong restriction on the possible form of the velocity fields \citep{groma2021dynamics}. The details are summarised below.

In this paper, our aim is to determine some of the parameters of the continuum theory of dislocations. In the first part of the paper the 3D continuum theory of curved dislocations is briefly summarised. Following that the consequences of the specific geometry, containing an impenetrable wall, considered are discussed. It is shown that for this case the continuum equations simplify to a form that is  numerically straightforwardly solvable.  In the last part of the paper this solution is directly compared to 2D and 3D DDD simulation results, from which three important parameters of the continuum theory can be determined.

\section{Problem setup}
Calibrating the parameters of the continuum theory is an important issue. To this end, the solutions of the continuum theory have to be compared to DDD simulation results. To be able to perform the comparison, a relatively simple but nontrivial problem is considered. Let us take a cubic simulation box with periodic boundary conditions and single slip geometry with Burgers vector $\vec{b}=(b,0,0)$. The box is orientated so that the slip plane is parallel to the $xy$ plane. To make the problem suitable for determining some of the parameters of the continuum model, let us introduce a narrow impenetrable wall in the simulation box that is parallel to the $yz$ plane. Since periodic boundary conditions are applied, the actual position of the wall does not make a difference.

After applying an external load, at equilibrium, the dislocation density does vary with the distance from the wall. As  explained below, this allows us to determine three parameters of the continuum theory. The problem considered is somewhat similar to the channel shearing problem discussed in Ref.~\cite{groma2003spatial}, but here we use periodic boundary conditions.

First, after a short summary of the continuum theory, the evolution equations of the statically  stored and GND densities are derived for this problem. Since the system has a $yz$ translation symmetry, the fields depend only on the distance from the wall ($x$) .

\section{Summary of the continuum theory of dislocations}
The details of the 3D continuum theory of curved dislocations in single slip can be found in the paper \cite{groma2021dynamics}. Here we provide only a short description of the theory.

The theory of curved dislocations in single slip is a direct generalisation of the 2D continuum theory of  straight parallel edge dislocations developed earlier through a systematic coarse-graining of the evolution equations of the individual dislocations \cite{groma1997link,groma2003spatial}. However, while the 2D continuum theory is directly linked to the discrete dislocation dynamics, building a direct link between the discrete and continuum descriptions for the curved dislocation problem seems virtually impossible. Therefore, in order to have a closed theory   one has to resort to phenomenological rules to express the dependence of the different dislocation velocities on the dislocation state \cite{groma2021dynamics}. However, the rules were deduced by closely following the 2D case 
\cite{groma2016dislocation,groma2018statistical,valdenaire2016density}.
\begin{figure}[pos=htbp]
\begin{center}
\begin{tcolorbox}[colframe=black!99!white,colback=white,width=(\linewidth-150pt)]
\color{red} Fields
\color{black}
\begin{equation*}
\rho(\bm{r},t),\ \kappa_1(\bm{r},t),\ \kappa_2(\bm{r},t),\ q(\bm{r},t)
\end{equation*}
\\ 
\color{red} Kinematics \color{black}
\begin{eqnarray*}
\partial_t \rho&=&-\partial_x(\rho v^\mathrm d_2)+\partial_y(\rho v^\mathrm d_1)+\partial_y(\kappa_1 v^\mathrm m)-\partial_x(\kappa_2 v^\mathrm m)
  \\ &&+q v^\mathrm m+ \lambda_1\rho\partial_y v^\mathrm d_1- \lambda_2 \rho\partial_x v^\mathrm d_2
  \\
 \partial_t \kappa_1&=&\partial_y(\rho v^\mathrm m+\kappa_1 v^\mathrm d_1+\kappa_2 v^\mathrm d_2)  \\
 \partial_t \kappa_2&=&-\partial_x(\rho v^\mathrm m+\kappa_1 v^\mathrm d_1+\kappa_2 v^\mathrm d_2)  \\
 \partial_t q &=&-\partial_x\left(q v^\mathrm d_2-v^\mathrm m Q_1\right)
 +\partial_y\left(q v^\mathrm d_1+v^\mathrm m Q_2\right)
\end{eqnarray*}
\color{red} Dynamics \color{black} \\ \vspace{8pt}
\centerline{Plastic potential: $P[\hat{\chi},\rho,\kappa_1,\kappa_2,q ]$ }\\ \color{red} \centerline{$\Downarrow$} \color{black} \\
\centerline{$\tau^*$, $\tau^\mathrm d_1$, $\tau^\mathrm d_2$} \vspace{5pt} \\ \vspace{10pt}
\color{red} \centerline{Velocities} \color{black} \\ \vspace{10pt}
\includegraphics[angle=0,width=7.5cm]{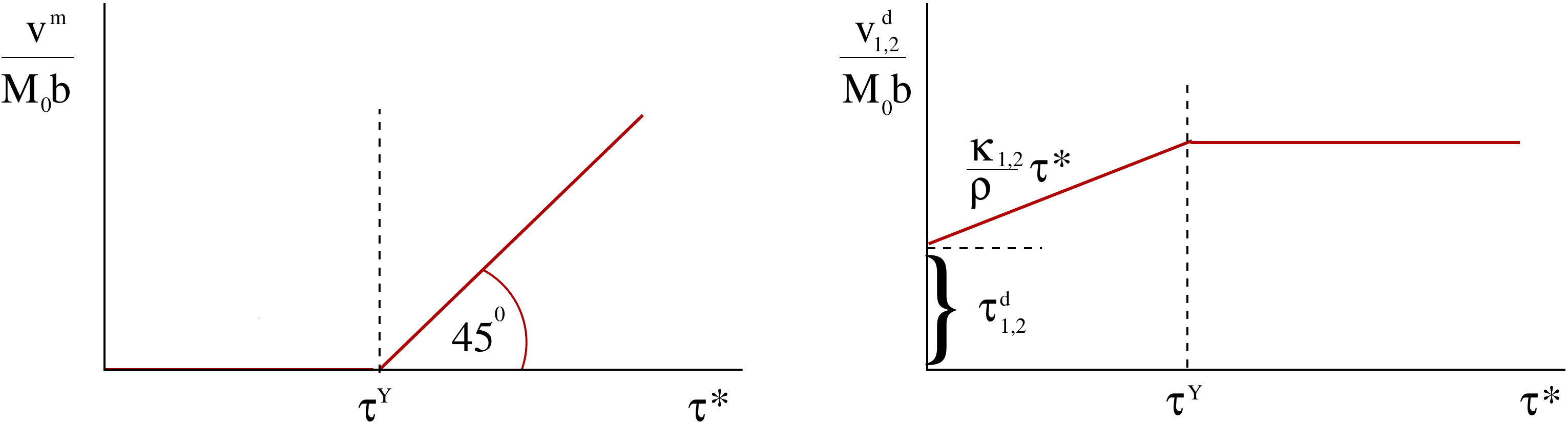}
\end{tcolorbox}
\end{center}
\caption{Summary of the model} \label{fig:sum}
\end{figure}

The proposed model  may be summarised as follows \citep{groma2021dynamics} (see also Fig.~\ref{fig:sum}) :
\begin{itemize}
 \item The state of the dislocation system is given by the fields: total dislocation density $\rho$, GND density vector $(\kappa_1,\kappa_2)$, and curvature density $q $.
 \item For the time evolution of these fields, a ``dipole'' approximation is used, meaning that in the Fourier expansion of the tangent angle dependence of the general fields $\rho'({\bf r},\varphi)$, $q'({\bf r},\varphi)$, and $v'({\bf r},\varphi)$, we stop at the first order.
 \item The dynamics of the system is obtained from a scalar functional $P[\hat{\chi},\rho,\kappa_1,\kappa_2,q ]$ called ``plastic potential'' (see in Refs. \cite{hochrainer2016thermodynamically,zaiser2015local, groma2021dynamics}. Its actual form is given below.  In analogy to irreversible thermodynamics, the relevant quantities are the  appropriate combinations of the spatial derivatives of the different ``chemical'' potentials, which are the corresponding functional derivatives of the plastic potential. The key quantities are the ``effective stress'' $\tau^*$ which is the sum of the mean field and ``back'' stresses, and the generalised ``diffusion'' stresses $\tau^\mathrm d_1$ and $\tau^\mathrm d_2$ which depend on the gradient of the dislocation density and the curvature field. For the actual dependence, see below.
\item The dependence of the velocity fields $v^\mathrm m$ and $v^\mathrm d_{1,2}$ on $\tau^*$ and $\tau^\mathrm d_{1,2}$ is indicated in Fig.~\ref{fig:sum}. Below flow stress, the mean velocity  $v^\mathrm m$ vanishes, while above it increases linearly with $\tau^*$. In the non-flowing regime the velocities  $v^\mathrm d_{1,2}$ are linear in $\tau^*$, whereas in the flowing regime, they remain constant upon increasing stress. The velocity relations suggested ensure that during the evolution of the system the plastic potential cannot increase.
\end{itemize}
Applying a somewhat different derivation method, Zhang \emph{et al.} obtained similar evolution equations \citep{zhang2025continuum}. It should also be noted that the dislocation continuum model shows a number of similarities with the strain-gradient plasticity models \citep{zhu1997strain,aifantis1999,fleck,gurtin}, but the key difference is that the length scale in the gradient terms is not a material parameter. It is the dislocation spacing, which is a spatially and temporally varying parameter i.e., it follows an evolution law. (Note, at $\rho\rightarrow 0$ the length scale goes to infinity. This case was addressed by \cite{schmitt2019mechanism}.)

\subsection{Stress calculation at small deformation limit}
As a first step, the mean-field stress has to be calculated. The method one should follow is summarised below \citep{groma2021dynamics}.  For small deformations, the elastic energy reads as
\begin{equation}
 E=\int \frac{1}{2}(\epsilon_{ij}-\epsilon_{ij}^p)C_{ijkl}(\epsilon_{lk}-\epsilon_{lk}^p)) dV, \label{eq:el}
\end{equation}
where
\begin{equation}
\epsilon_{ij}=\frac{1}{2} \left(\partial_i u_j+\partial_j u_i\right)
\end{equation}
is the total deformation, $u_i$ is the displacement field,  $\epsilon_{ij}^p$ is the plastic deformation, and $C_{ijkl}$ is the elastic modulus tensor. According to Kröner \citep{kroner1981continuum} the equilibrium equation corresponding to the energy given by   Eq. (\ref{eq:el}) reads as
\begin{equation}
 \partial_i C_{ijkl}(\epsilon_{lk}-\epsilon_{lk}^p)=0
\end{equation}
and the stress is
\begin{equation}
 \sigma_{ij}=C_{ijkl}(\epsilon_{lk}-\epsilon_{lk}^p).
\end{equation}
For further consideration, we introduce the incompatibility operator acting on a matrix $A_{ij}$
\begin{equation}
 (Inc(\hat{A})_{ij}=e_{ikm}e_{jln}\partial_k\partial_l A_{mn}.
\end{equation}
It is easy to see that \citep{kroner1981continuum}
\begin{equation}
 Inc(\hat{\epsilon})=0
\end{equation}
and
\begin{equation}
 \partial_i Inc(\hat{A})_{ij}=0 \label{eq:divInc}
\end{equation}
for any $\hat{A}$.

Since the divergence of the stress vanishes, it is useful to introduce the stress potential $\varPsi_{mn}$,  the incompatibility of which is the stress \citep{kroner1981continuum}
\begin{equation}
 (Inc\hat{\varPsi})_{ij}=\sigma_{ij}.
\end{equation}
Due to the identity (\ref{eq:divInc}) the stress equilibrium condition is automatically fulfilled.

In order to directly calculate the stress state generated by the dislocation, let us consider the functional (called the mean field plastic potential) \citep{groma2007dynamics,groma2010variational,groma2016dislocation}
\begin{equation}
 P^{mf}[\varPsi_{ij},\eta_{ij}]=\int \left[ \frac{1}{2} (Inc\hat{\varPsi})_{ij}
 C_{ijlk}^{-1}(Inc\hat{\varPsi})_{kl}+\eta_{ij}\varPsi_{ji}
 \right]dV, \label{eq:Pmf}
\end{equation}
where
\begin{equation}
 \eta_{ij}=(Inc \epsilon^p)_{ij} \label{eq:eta}
\end{equation}
is the incompatibility tensor \citep{kroner1981continuum}. The stress potential at the minimum of $P^{mf}$ satisfies the equation \citep{groma2010variational}
\begin{equation}
 \frac{\delta P^{mf}}{\delta \varPsi_{ij}}=
 Inc(\hat{C}^{-1} Inc\hat{\varPhi})_{ij}+\eta_{ij}=0.
 \label{eq:minphi}
\end{equation}
However, since
\begin{equation}
 \epsilon^e_{ij}=C^{-1}_{ijkl}\sigma_{ik}
 =C^{-1}_{ijkl}(Inc \hat{\varPsi})_{ik}
\end{equation}
and
\begin{equation}
 Inc(\hat{\epsilon}^e)=Inc(\hat{\epsilon}-\hat{\epsilon}^p)=
 -Inc(\hat{\epsilon}^p)
\end{equation}
with $\epsilon^e_{ij}=\epsilon_{ij}-\epsilon^p_ {ij}$ denoting the elastic deformation (\citep{kroner1981continuum}).
 Eq. (\ref{eq:minphi}) ensures Eq. (\ref{eq:eta}).
Thus, one can determine the stress state generated by dislocations by solving Eq.~(\ref{eq:minphi}) for $\varPsi_{ij}$ and taking the incompatibility of the solution.
It should be noted that $\varPsi_{ij}$ is not uniquely defined, there is a  gauge freedom, but certainly the stress is uniquely defined
\citep{kroner1981continuum}.

\section{Stress calculation for the wall problem}
For the specific ``wall'' problem considered in the paper, the first issue we should discuss is the calculation of the mean-field shear stress generated by the dislocation ensemble considered. As explained above, first, the incompatibility tensor
\begin{equation}
 \eta_{ij}=e_{ikm}e_{jln}\partial_k\partial_l \epsilon^p_{mn}
\end{equation}
has to be calculated. Assuming that the Burgers vector is parallel to the $x$ axis $\vec{b}=(b,0,0)$, for the geometry considered  only the $\beta^p_{31}$ component of the plastic distortion is different from zero. For  brevity, the plastic shear $\gamma^p=\beta^p_{31}$ is introduced.  In this case, the dislocation density tensor $\alpha_{ij}=e_{ikl}\partial_k \beta^p_{lj} $ is
\begin{equation} \hat{\alpha}=b
 \left( \begin{array}{ccc}
          \kappa_1 & \kappa_2 & 0 \\
          0 & 0 & 0 \\
          0 & 0 & 0
        \end{array}
\right), \label{eq.alpha}
\end{equation}
where
\begin{eqnarray}
  \kappa_1&=&\frac{1}{b}\partial_2 \gamma^p \label{eq.kappa1}\\
  \kappa_2&=&-\frac{1}{b}\partial_1 \gamma^p. \label{eq.kappa2}
\end{eqnarray}
An important consequence of the above form is that
\begin{equation}
 \partial_1 \kappa_1 + \partial_2 \kappa_2=0 \label{eq.divkappa}
 \end{equation}
which ensures that $\partial_i \alpha_{ij}=0$.  Applying the relation (derived in Ref. \cite{kroner1981continuum})
\begin{equation}
\eta_{ij}=-\frac{1}{2}(e_{iln}\partial_n \alpha_{jl}+
e_{jln}\partial_n\alpha_{il})
\end{equation}
for $\hat{\alpha}$ given by Eq.(\ref{eq.alpha}) one obtains that
\begin{equation} \hat{\eta}=\frac{b}{2}
 \left( \begin{array}{ccc}
          0 & -\partial_3\kappa_1 & \partial_2 \kappa_1 \\
          -\partial_3 \kappa_1 & -2\partial_3 \kappa_2 & \partial_2 \kappa_2 \\
          \partial_2 \kappa_1 & \partial_2 \kappa_2 & 0
        \end{array}
\right). \label{eq.eta}
\end{equation}
It should be noted that due to Eq. (\ref{eq.divkappa}) $\eta_{32}=\eta_{23}=\partial_2\kappa_2=-\partial_1\kappa_1$. As a result, as expected, $\partial_i \eta_{ij}=0$.

The stress tensor $\sigma_{ij}$ can be given as the incompatibility of the stress potential $\varPsi_{ij}$ \citep{kroner1981continuum}, i.e.
\begin{equation}
 \sigma_{ij}=-e_{ikm}e_{jln}\partial_k\partial_l \varPsi_{mn}. \label{eq.sigma}
\end{equation}
For isotropic materials, it is useful to introduce the quantity
\begin{equation}
 \varPsi'_{ij}=\frac{1}{2\mu}\left(\varPsi_{ij}-\frac{\nu}{1+2\nu} \varPsi_{ll}\delta_{ij}\right).\label{eq.psi'}
\end{equation}
It can be seen that  this new variable fulfils the biharmonic equation (see \cite{kroner1981continuum})
\begin{equation}
 \bigtriangleup^2 \varPsi'_{ij}=\eta_{ij},
\end{equation}
following that
\begin{equation}
  \varPsi'_{ij}=\bigtriangleup^{-2}\eta_{ij}, \label{eq.solution}
\end{equation}
where $\bigtriangleup^{-2}$ denotes the inverse of the operator $\bigtriangleup^{2}$.
In our case from Eqs. (\ref{eq.eta}, \ref{eq.psi'},\ref{eq.psi}) one can find that
\begin{equation} \hat{\varPsi}'=
 \left( \begin{array}{ccc}
          0 & \varPsi'_{12} & \varPsi'_{13}   \\
          \varPsi'_{12}  & \varPsi'_{22}  & \varPsi'_{23}  \\
          \varPsi'_{13}  & \varPsi'_{23}  & 0
        \end{array}
\right). \label{eq.Psit}
\end{equation}
After inverting the relation (\ref{eq.psi'}) one obtains that
\begin{equation}
 \varPsi_{ij}=2\mu\left(\varPsi'_{ij}+\frac{\nu}{1-\nu} \varPsi'_{ll}\delta_{ij}\right).\label{eq.psi}
\end{equation}
For our further considerations, we need the shear stress $\sigma_{13}$. From Eqs. (\ref{eq.sigma},\ref{eq.Psit})
\begin{equation}
 \sigma_{13}=\partial_1(\partial_2\varPsi_{23}-\partial_3\varPsi_{22})-\partial_2(\partial_2\varPsi_{13}-\partial_3\varPsi_{12}),
\end{equation}
or with $\varPsi'_{ij}$
\begin{equation}
 \sigma_{13}=2\mu\left[\partial_1\partial_2\varPsi'_{23}-\frac{1}{1-\nu}\partial_1\partial_3\varPsi'_{22}-\partial_2\partial_2\varPsi'_{13}+\partial_2\partial_3\varPsi'_{12}\right].
\end{equation}
After substituting expression (\ref{eq.solution}) into the above equation, we get
\begin{equation}
 \sigma_{13}=b\mu\bigtriangleup^{-2} \left[\partial_1\partial_2\partial_2\kappa_2+\frac{1}{1-\nu}\partial_1\partial_3\partial_3\kappa_2-\partial_2\partial_2\partial_2\kappa_1-\partial_2\partial_3\partial_3\kappa_1\right],
\end{equation}
or with $\gamma^p$
\begin{equation}
 \sigma_{13}=-\mu\bigtriangleup^{-2} \left[\partial_1\partial_1\partial_2\partial_2+\frac{1}{1-\nu}\partial_1\partial_1\partial_3\partial_3+\partial_2\partial_2\partial_2\partial_2+\partial_2\partial_2\partial_3\partial_3\right]\gamma^p. \label{eq.stress}
\end{equation}

\section{The evolution equations}

As obtained in Ref.~\cite{groma2021dynamics}, the time evolution of the fields $\rho, \kappa_1, \kappa_2$ and $q$ is described by the following equations:
\begin{eqnarray}
 \partial_t \kappa_1&=&\partial_2(\rho v^\mathrm m+\kappa_1 v^\mathrm d_1+\kappa_2 v^\mathrm d_2)
 \label{eq.kappa1t} \\
 \partial_t \kappa_2&=&-\partial_1(\rho v^\mathrm m+\kappa_1 v^\mathrm d_1+\kappa_2 v^\mathrm d_2) \label{eq.kappa2t}
 \end{eqnarray}
\begin{eqnarray}
 \partial_t \rho&=&-\partial_1(\rho v^\mathrm d_2)+\partial_2(\rho v^\mathrm d_1)+\partial_2(\kappa_1 v^\mathrm m)-\partial_1(\kappa_2 v^\mathrm m)
 +q v^\mathrm m+ \lambda_1\rho\partial_2 v^\mathrm d_1- \lambda_2 \rho\partial_1 v^\mathrm d_2 \label{eq.rhot}
 \end{eqnarray}
 \begin{eqnarray}
 \partial_t q &=&-\partial_1\left(q v^\mathrm d_2-v^\mathrm m Q_1\right)
 +\partial_2\left(q v^\mathrm d_1+v^\mathrm m Q_2\right),
 \label{eq.qt}
\end{eqnarray}
where $v^m$ and $v_{1,2}^d$ are the mean and ``drift'' velocities, respectively. They are functions of the fields $\rho, \kappa_1, \kappa_2$ and $q$. For their detailed form, see Ref.~\cite{groma2021dynamics}. Moreover,
\begin{equation}
 Q_i=\partial_i \rho \label{eq.Q}
\end{equation}
and $\lambda_{1,2}$ are parameters that depend on the curvature. For small curvature $q$ $\lambda_{1,2}$  are proportional to $q^2/\rho^3$, so keeping only the linear terms in $q$, we are going to consider, they can be neglected.
It should be noted that since $\kappa_1=\partial_2 \gamma^p/b$ and $\kappa_2=-\partial_1 \gamma^p/b$, by  appropriate derivations, Eqs. (\ref{eq.kappa1t},\ref{eq.kappa2t}) can be obtained from the equation
\begin{eqnarray}
 \partial_t \gamma^p=b\left(\rho v^\mathrm m+\kappa_1 v^\mathrm d_1+\kappa_2 v^\mathrm d_2\right),
 \label{eq.betat}
 \end{eqnarray}
that can be called as a ``generalised Orowan's law''.

To be able to obtain the actual forms of the two velocities, the form of the plastic potential $P[\varPsi_{ij},\eta_{ij},\rho,\kappa_1,\kappa_2,q]$ has to be given. As discussed in Ref.~\cite{groma2021dynamics}, it is the sum of two terms, the $P^{mf}[\varPsi_{ij},\eta_{ij}]$ mean field term  given by Eq. (\ref{eq:Pmf}) and the $P^{corr}[\rho,\kappa_1,\kappa_2,q]$ term with the form
\begin{equation}
 P^{corr}=\int \frac{\mu}{2\pi(1-\nu)} b^2 \left[A \rho \ln\left(\frac{\rho}{\rho_0}\right)+
 \frac{\bm\kappa \cdot \hat{ D}  \cdot \bm \kappa }{2\rho}+\rho R\left(\frac{q ^2}{\rho^3}\right) \right] \mathrm dV,
 \label{eq:pcorr1}
\end{equation}
where  $A$  is a dimensionless constant, $\hat{D}$ is a $2 \times 2$ dimensionless constant matrix \citep{groma2021dynamics}, and $\rho_0=1/c^2b^2\gg \rho$ is a constant parameter with dislocation density dimension with $c$ being a constant determined by the core properties of the dislocations \cite{groma2016dislocation,groma2021dynamics}. The last term on the right hand side of the above expression accounts for the energy related to dislocation curvature. Since in most cases the radius of curvature of a dislocation is much larger than the dislocation spacing, the dimensionless quantity $q^2/\rho^3$ is small. Thus, the function $R(q^2/\rho^3)$ can be well approximated with a linear function  $R(q^2/\rho^3)=R_0 q^2/\rho^3$.

For the problem considered here, we need only the functional form of $v^m$ and $v_{1,2}^d$ in the flowing regime \citep{groma2021dynamics} where the effective shear stress 
\begin{equation}
    \tau^* = \tau^\mathrm{mf} + \tau^\mathrm b=\tau_0+\sigma_{13}+\tau^\mathrm b
\end{equation}
is larger than the flow stress
\begin{equation}
 \tau^Y=\alpha \mu b \sqrt{\rho}.
\end{equation}
Here  $\tau_0$ is the external shear stress applied and $\tau^\mathrm b$ is the back-stress, that is,
\begin{equation}
 \tau^\mathrm b=\frac{\mu b}{2\pi(1-\nu)\rho}\left[\partial_2\left(D_{11}\kappa_1+D_{12}\kappa_2\right)
 -\partial_1\left(D_{22}\kappa_2+D_{12}\kappa_1\right)\right].
\end{equation}
With $\gamma^p$ the above expression reads as
\begin{equation}
 \tau^\mathrm b=\frac{\mu }{2\pi(1-\nu)\rho}\left[D_{11}\partial_2\partial_2 +D_{22}\partial_1\partial_1-(D_{12}+D_{21})\partial_1\partial_2\right]\gamma^p.
\end{equation}
In the flowing regime \citep{groma2021dynamics} (see Fig. 1)
\begin{equation}
 v^\mathrm m=M_0b(\tau^*-\alpha \mu b \sqrt{\rho})
 =M_0b(\tau_0+\sigma_{13}+\tau^\mathrm b-\alpha \mu b \sqrt{\rho})
\end{equation}
and
\begin{equation}
 v^\mathrm d_{i}=M_0b\left(\frac{\kappa_i}{\rho} \alpha \mu b \sqrt{\rho}+\tau^\mathrm d_i\right)
 \ \ i=1,2,
\end{equation}
with
\begin{equation}
 \tau^\mathrm d_1=\frac{\mu b}{2 \pi (1-\nu)\rho} \left[ A \left(1+2\lambda_1+\lambda_1 \ln \frac{\rho}{\rho_0} \right) \partial_2 \rho + R_0 q \, \partial_2 \frac{q}{\rho^2} \right]
 \label{eq:tau_1}
\end{equation}
\begin{equation}
 \tau^\mathrm d_2=-\frac{\mu b}{2 \pi (1-\nu)\rho} \left[ A \left(1+2\lambda_2+\lambda_2 \ln \frac{\rho}{\rho_0} \right) \partial_1 \rho + R_0 q \, \partial_1 \frac{q}{\rho^2} \right].
 \label{eq:tau_2}
\end{equation}
As mentioned earlier, for the problem considered, the term proportional to $\lambda_{1,2}$ can be neglected (they are quadratic in $q$).
Moreover, for the same reason the second terms on the right hand side of the above equations can also be neglected. Thus,
\begin{equation}
 \tau^\mathrm d_1=\frac{\mu b A}{2 \pi (1-\nu)\rho}    \partial_2 \rho
\end{equation}
\begin{equation}
 \tau^\mathrm d_2=-\frac{\mu b A}{2 \pi (1-\nu)\rho}  \partial_1 \rho.
\end{equation}
With these
\begin{equation}
 v^\mathrm d_{1}=M_0b^2\mu\left(\frac{\kappa_1}{\rho} \alpha\sqrt{\rho}+\frac{A^*}{\rho}\partial_2 \rho\right)
\end{equation}
\begin{equation}
 v^\mathrm d_{2}=M_0b^2\mu\left(\frac{\kappa_2}{\rho} \alpha \sqrt{\rho}-\frac{A^*}{\rho}\partial_1 \rho\right),
\end{equation}
where $A^*=A/2 \pi (1-\nu)$ is introduced for a shorter notation.

\section{The evolution equations for the problem considered}
According to the considerations discussed in the previous section, the evolution equations in the ``flowing'' regime read as follows.
\begin{eqnarray}
 \partial_t \gamma_p=b\rho v^\mathrm m+v^\mathrm d_1\partial_2\gamma_p - v^\mathrm d_2 \partial_1 \gamma_p
 \label{eq.betat2}
 \end{eqnarray}
\begin{eqnarray}
 \partial_t \rho&=&-\partial_1(\rho v^\mathrm d_2)+\partial_2(\rho v^\mathrm d_1)+\partial_2(bv^\mathrm m\partial_2 \gamma_p)+\partial_1(bv^\mathrm m\partial_1 \gamma_p)
 +q v^\mathrm m\label{eq.rhot2}
 \end{eqnarray}
 \begin{eqnarray}
 \partial_t q &=&-\partial_1\left(q v^\mathrm d_2-v^\mathrm m \partial_1 \rho\right)
 +\partial_2\left(q v^\mathrm d_1+v^\mathrm m \partial_2 \rho\right)
 \label{eq.qt2}
\end{eqnarray}
\begin{equation}
 v^\mathrm m=M_0b(\tau_0+\sigma_{13}+\tau^\mathrm b-\alpha \mu b \sqrt{\rho})
\end{equation}
\begin{equation}
 v^\mathrm d_{1}=M_0\frac{b^2\mu}{\rho}\left( \alpha \sqrt{\rho}\partial_2 \frac{\gamma^p}{b}+A^*\partial_2 \rho\right)
\end{equation}
\begin{equation}
 v^\mathrm d_{2}=-M_0\frac{b^2\mu}{\rho}\left(\alpha \sqrt{\rho}\partial_1 \frac{\gamma^p}{b}+A^*\partial_1 \rho\right).
\end{equation}

Let us consider a problem in which the dislocation distributions vary only in the $x$ direction (the variations in the $y$ and $z$ directions are averaged out). In this case, the above equations simplify to:
\begin{eqnarray}
 \partial_t \beta_0=\rho v^\mathrm m - v^\mathrm d_2 \partial_1 \beta_0
 \label{eq.betat3}
 \end{eqnarray}
\begin{eqnarray}
 \partial_t \rho&=&-\partial_1(\rho v^\mathrm d_2)+\partial_1(v^\mathrm m\partial_1 \beta_0)
 +q v^\mathrm m\label{eq.rhot3}
 \end{eqnarray}
 \begin{eqnarray}
 \partial_t q &=&-\partial_1\left(q v^\mathrm d_2-v^\mathrm m \partial_1 \rho\right)
 \label{eq.qt3}
\end{eqnarray}
\begin{equation}
 v^\mathrm m=M_0b(\tau_0+\sigma_{13}+\tau^\mathrm b-\alpha \mu b \sqrt{\rho})
\end{equation}
\begin{equation}
 v^\mathrm d_{2}=-M_0\frac{b^2\mu}{\rho}\left(\alpha \sqrt{\rho}\partial_1 \frac{\gamma_p}{b}+A^*\partial_1 \rho\right),
\end{equation}
where the back stress $\tau^b$ is
\begin{eqnarray}
 \tau^\mathrm b=\frac{\mu }{2\pi(1-\nu)\rho}D_{22}\partial_1\partial_1\gamma_p,
\end{eqnarray}
As seen from Eq. (\ref{eq.stress}), the varying $\gamma^p$ does not generate stress in the case considered here, i.e. $\sigma_{13}=0$. So, the shear stress appearing in the equations is simply the external shear.

Let us consider the stationary state or nearly stationary state. This means that the time derivatives of the different quantities appearing in the above equations can be neglected. Thus,
\begin{eqnarray}
 \rho v^\mathrm m + v^\mathrm d_2 \kappa_2=0
 \label{eq.betat3s}
 \end{eqnarray}
\begin{eqnarray}
 \partial_1(\rho v^\mathrm d_2+v^\mathrm m\kappa_2)
 =q v^\mathrm m \label{eq.rhot3s}
 \end{eqnarray}
 \begin{eqnarray}
 \partial_1\left(q v^\mathrm d_2-v^\mathrm m \partial_1 \rho\right)=0
 \label{eq.qt3s}
\end{eqnarray}
\begin{equation}
 v^\mathrm m=M_0b(\tau_0+\tau^\mathrm b-\alpha \mu b \sqrt{\rho})
\end{equation}
\begin{equation}
 v^\mathrm d_{2}=-M_0\frac{b^2\mu}{\rho}\left(-\alpha \sqrt{\rho}\kappa_2+A^*\partial_1 \rho\right),
\end{equation}
with
\begin{eqnarray}
 \tau^\mathrm b=-\frac{\mu b}{2\pi(1-\nu)\rho}D_{22}\partial_1\kappa_2,
\end{eqnarray}
where the relation $\kappa_2=-\frac{1}{b}\partial_1\gamma^p$ was reintroduced into the equations.

In the following we are going to consider only the steady state configuration. In this case, the two velocities have to vanish. Since in the DDD simulations we reach the steady state from an initially everywhere ``flowing'' state, it can be assumed that the spatial evolution of the system stops everywhere when $\tau^*$ equals the local yield stress. Thus, from Eqs. (\ref{eq.betat3s}, \ref{eq.rhot3s}, \ref{eq.qt3s}) $v^m$ and $v_2^d$ vanish if
\begin{equation}
 \tau_0-\alpha \mu b \sqrt{\rho}=\frac{\mu b}{2\pi(1-\nu)\rho}D_{22}\partial_1\kappa_2
 \label{eq.f}
\end{equation}
\begin{equation}
 \alpha \sqrt{\rho}\kappa_2=A^*\partial_1\rho. \label{eq:kappa}
\end{equation}
The two equations given above make it possible to determine  the parameters $D_{22}$, $A^*$ and $\alpha$ from a series of DDD simulations corresponding to different external shear $\tau_0$. However, since they contain the spatial derivative of the fields to reduce the numerical noise that one would experience after numerical derivation, a method summarised below is proposed.

It should be noted in earlier publications (see \citep{groma2003spatial}) it was assumed, that next to the wall, the dislocations predominantly have the same  sign i.e. $\kappa_2=\pm \rho$ and the flow stress is negligible beside the external load. In this case, Eq.(\ref{eq.f}) predicts a nearly exponential decay. However, with the results explained above, a more refined analysis is possible.

\begin{figure}[pos=htbp]
 \centerline{\includegraphics[angle=0,width=6cm]{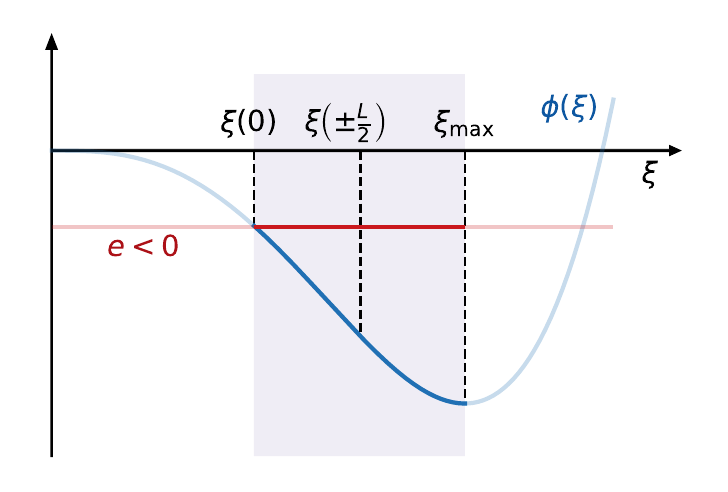}}
 \caption{The effective potential $\phi$}
 \label{fig:Phi}
\end{figure}

By combining the above two equations, we can obtain that
\begin{equation}
 \tau_0 \left(\sqrt{\rho}\right)^2-\alpha \mu b \left(\sqrt{\rho}\right)^3=\frac{\mu b}{\pi(1-\nu)\alpha}A^* D_{22} \partial_1 \partial_1 \sqrt{\rho}.
\end{equation}
If we introduce the notation $\xi(x)=\sqrt{\rho}$ and multiply the above equation by $\partial_1 \xi$ we get
\begin{equation}
 \partial_1\left\{\frac{\tau_0}{3} \xi^3-\frac{\alpha \mu b}{4} \xi^4\right\}=\partial_1\left\{ \frac{\mu b}{\pi(1-\nu)\alpha}\frac{A^* D_{22}}{2}\left(\partial_1 \xi\right)^2\right\}. \label{eq:work}
\end{equation}
For shorter notation, it is useful to introduce the quantities
\begin{equation}
 \xi_{max}=\frac{\tau_0}{\alpha \mu b} \label{eq:xi_max}
\end{equation}
and
\begin{equation}
 C=\sqrt{\frac{2\pi(1-\nu)\alpha^2}{A^* D_{22}}}. \label{eq:C}
\end{equation}
Actually, $\rho_{max}=\xi^2_{max}$ is the maximum  dislocation density that can be in the system at $\tau_0$ external load assuming we are in the flowing regime everywhere in the system. With this, Eq. (\ref{eq:xi_max}) reads as
\begin{equation}
 \partial_1\left\{\xi_{max}\frac{ \xi^3}{3} -\frac{\xi^4}{4} \right\}=\partial_1\left\{ \frac{1}{C^2}\left(\partial_1 \xi\right)^2\right\}. \label{eq:work2}
\end{equation}
This means that the quantity
\begin{equation}
 e=  \frac{1}{C^2}\left(\partial_1 \xi\right)^2-\left\{\xi_{max}\frac{\xi^3}{3}-\frac{\xi^4}{4}\right\}
\end{equation}
is conserved in space. If we extract $\partial_1\xi$ from the equation we obtain that
\begin{equation}
 \partial_1 \xi=\pm C \sqrt{e+\xi_{max}\frac{\xi^3}{3}-\frac{\xi^4}{4}}. \label{eq:xit}
\end{equation}
The quantity
\begin{equation}
 \phi(\xi)=-\xi_{max}\frac{\xi^3}{3}+\frac{\xi^4}{4}
\end{equation}
acts as an ``effective'' potential. Due to symmetry reason at the edge of our simulation box (say $x=0$)  the $\partial_1 \xi$ should vanish. Thus,
\begin{equation}
 e=\phi(\sqrt{\rho(x=0)}.
\end{equation}
Since the minimum of the potential is at the condition $\xi=\xi_{max}$ that, according to Eq. (\ref{eq:xi_max}), corresponds to $\tau_0=\alpha \mu b \sqrt{\rho_{max}}$, if we are in the flowing regime the dislocation density at $x=0$ has to be smaller than the one corresponding to the minimum. This follows that
$e<0$. Then the equation $e=\phi(\xi_1)$ has another solution that is positive. The minimum of $\phi(\xi)$ is between $\xi_0$ and $\xi_1$. Mathematically, the possible values of $\xi$ values have to be in between $\xi_0$ and $\xi_1$ but since during the evaluation of the system we are in the flowing regime $\xi$ cannot be larger than $\xi_{max}=\sqrt{\rho_{max}}$. Thus, $\xi$ is in the interval $(\xi_0,\xi_{max})$. So, $\xi(L/2)$ corresponding to the dislocation density at the wall is also within this interval  (see Fig. \ref{fig:Phi}).
From Eq. (\ref{eq:xit}), one obtains that
\begin{equation}
 \frac{1}{C}\int_{\xi(0)}^{\xi(x)} \frac{1}{\sqrt{e+\xi_{max}\frac{\xi^3}{3}-\frac{\xi^4}{4}}} d\xi=x. \label{eq.ff}
\end{equation}
Now $C$ should be such that the integral on the left hand side at $\xi=\xi(x)$ gives precisely $x$. By fitting the function $x(\xi,C, \alpha)$ to the pairs $(x_i,\sqrt{\rho_i})$ obtained by DDD simulation, the parameters $C$ and $\alpha$ can be determined.

Since $C$ depends on $A^*$ and $D_{22}$, to determine separately the two correlation parameters  Eq. (\ref{eq:kappa}) should also be used. Due to the fact that numerical differentiation can enhance noise, it is useful to integrate Eq. (\ref{eq:kappa})  with respect to $x$ giving
\begin{equation}
 \int_{x_0}^x \kappa_2 dx=\frac{A^*}{\alpha}(\xi(x)-\xi(x_0)).
 \label{eq:intkappa}
\end{equation}
This follows that plotting the integral of $\kappa_2$ versus  $\xi$
should be a straight line. The slope of the line gives $A^*/\alpha$

It should be noted that on the two sides of the wall, the sign of GND is different. Thus, next to the wall, the GND varies rapidly. Therefore, the derivative of $\kappa_2$ with respect to $x$ is large near the wall. Since in the considerations discussed above small spatial derivatives were assumed, one cannot expect that Eq.  (\ref{eq.ff}) describes the spatial variation of the fields close to the wall. For that, higher derivatives in the plastic potential should be introduced (see Ref.~\cite{groma2015scale}). As a consequence, to determine the numerical parameters, an appropriate region near to the wall has to be excluded.

\section{Numerical results}

Since Eqs. (\ref{eq.ff},\ref{eq:intkappa}) also describe the spatial variation of $\rho$ and $\kappa$  for the problem of a straight parallel edge dislocation system with single slip \citep{groma2003spatial}, the results of both 2D and 3D DDD simulations were analysed to determine the three parameters of the corresponding continuum theory of dislocations.
Certainly, the actual values of the parameters are not expected to be exactly the same.

\subsection{The 2D case}
In the 2D case, for the direct comparison of the theoretical results with numerical simulations, it is useful to introduce the following notations:
\begin{equation}
 \tau'=\frac{1}{\sqrt{<\rho >}}\frac{2\pi(1-\nu)}{\mu b}\tau_0,
\end{equation}
\begin{equation}
 \xi'=\frac{\xi}{\sqrt{<\rho >}}=\sqrt{\frac{\rho}{<\rho>}},
\end{equation}
\begin{equation}
 \xi'_{max}=\frac{\xi_{max}}{\sqrt{<\rho >}}=\frac{\tau_0}{\alpha \mu b \sqrt{<\rho >}}=\frac{\tau'}{2\pi(1-\nu)\alpha},
 \label{eq:2D_xi_max}
\end{equation}
\begin{equation}
 \kappa'=\frac{\kappa_2}{<\rho>},
\end{equation}
and
\begin{equation}
 x'=\frac{x}{L},
\end{equation}
where $<\rho >$ is the mean dislocation density in the system and $L$ is the system size.
With these notations, Eqs. (\ref{eq.ff},\ref{eq:intkappa}) read as
\begin{equation}
 \phi(\xi')=<\rho >^2\left(-\frac{\xi'_{max}}{3} \xi'^3+\frac{1}{4} \xi'^4)\right)
\end{equation}
\begin{equation}
 \int_{x'_0}^{x'} \kappa' dx'=\frac{A^*}{\sqrt{N}\alpha}(\xi'(x)-\xi'(x_0)).
 \label{eq:intkappa'}
\end{equation}
Thus, Eq. (\ref{eq.ff}) reads as
\begin{equation}
 \frac{1}{C'}\int_{\xi'(0)}^{\xi'(x')} \frac{1}{\sqrt{e'+\frac{\xi'_{max}}{3} \xi'^3-\frac{1}{4} \xi'^4}} d\xi'=x' \label{eq.f2}
\end{equation}
with
\begin{equation}
 e'=\left.\frac{\xi'_{max}}{3} \xi'^3-\frac{1}{4} \xi'^4\right|_{\xi'(0)}
\end{equation}
and
\begin{equation}
 C'^2=C^2 L^2<\rho>=C^2 N,
 \end{equation}
where $N=L^2<\rho>$ is the number of dislocations in the system, which is constant in the 2D case.

\begin{figure}[pos=htbp]
 \centerline{\includegraphics[height=8cm]{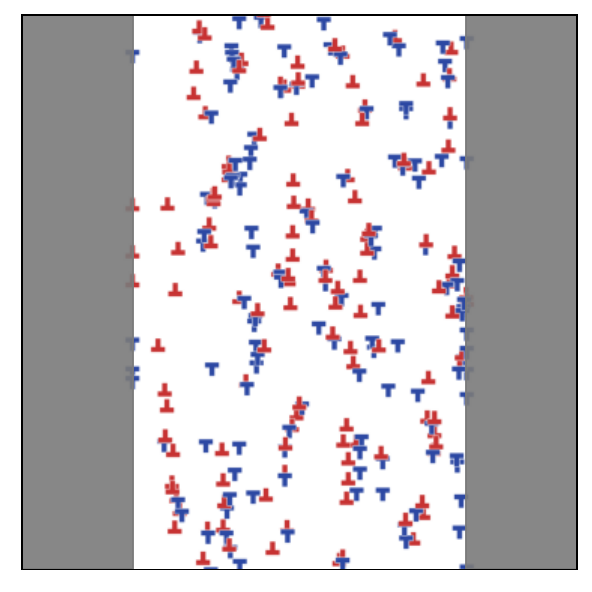}}
 \caption{Typical 2D dislocation configuration with periodic boundary conditions at $\tau'=12$. The gray area is not penetrable for the dislocations.} \label{fig:2D_2}
\end{figure}

\begin{figure}[pos=htbp]
 \includegraphics[height=8cm]{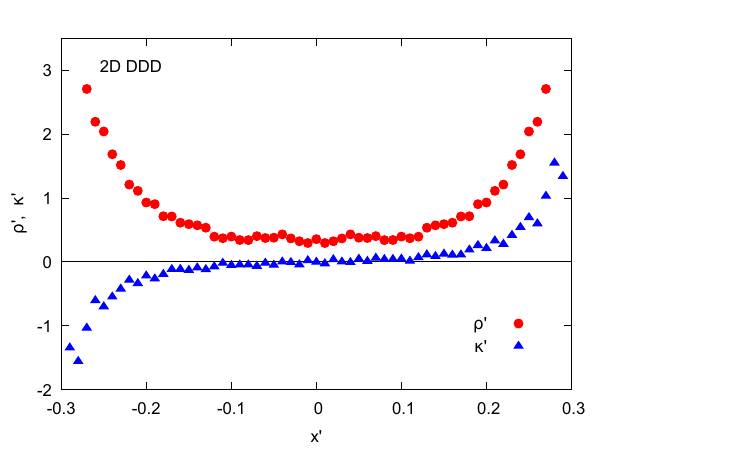}
 \caption{The statistically stored and the GND densities of dislocations as a function of the distance from the impenetrable zone. $\tau'=12$.} \label{fig:2D_rho}
\end{figure}

\begin{figure}[pos=htbp]
 \includegraphics[height=8cm]{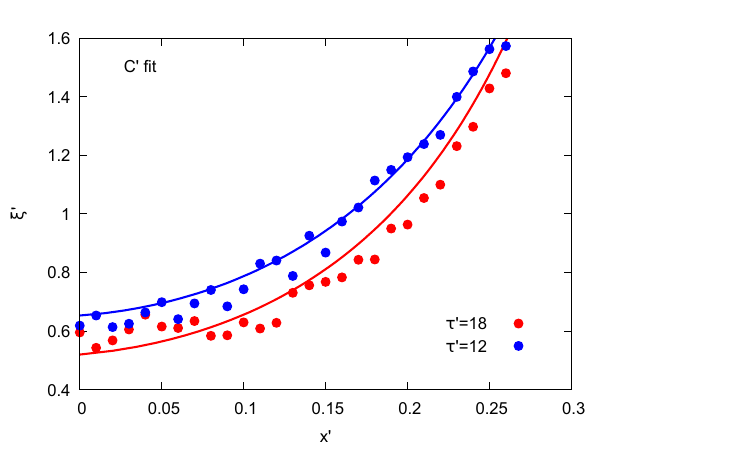}
 \caption{2D simulation results.  The $\xi'$ versus $x'$ plots at two different external loads.} \label{fig:2Dfit}
\end{figure}

\begin{figure}[pos=htbp]
 \includegraphics[height=8cm]{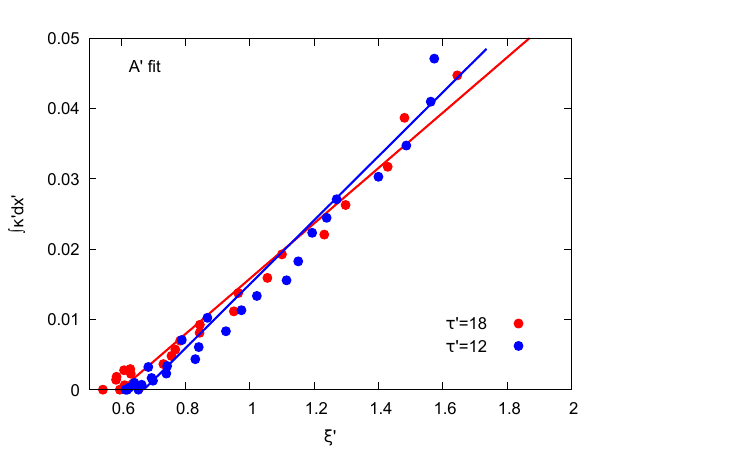}
 \caption{2D simulation results.  The $\int \kappa dx$ versus $\xi'$ plots at two different external loads.} \label{fig:2Dkappafit}
\end{figure}

The 2D simulations were performed with an efficient implicit method (details are given in Ref.~\cite{peterffy2020efficient}).  A typical dislocation configuration obtained  is seen in Fig. \ref{fig:2D_2} at $\tau'=12$. The simulations were performed with periodic boundary conditions, and the dislocation mobility was set to zero within the grey area. The $x$ size of the active area is $0.6L$. The total number of dislocations is 256, and 100 simulations were performed with different initial dislocation configurations. At the beginning of the simulations, the dislocation system was relaxed and then an external load was applied.
The functions $\rho(x)$ and $\kappa_2(x)$ obtained by averaging the 100 simulations are shown in Fig. \ref{fig:2D_rho}. The $x$ coordinate is measured from the middle of the ``active'' area.  As expected, due to the external load, on the two sides of the wall layers develop with enhanced   $\rho(x)$ and $\kappa_2(x)$.
The curves $\xi'$ versus $x'$ with the fitted theoretical functions at two different external stress levels  are plotted in Fig. \ref{fig:2Dfit}.
The curves $\int \kappa' dx$ versus $\xi'$ for the two applied load levels are plotted in Fig. \ref{fig:2Dkappafit}.

In the evaluation, the data sets corresponding to the two stress levels were analysed simultaneously.  The actual values of the parameters obtained are: $\alpha= 0.3$, $C'= 3\pm 0.1 $ and $A^*/(\sqrt{N}\alpha)= 0.04\pm 0.002 $. With $N=256$, one gets $C=0.19\pm 0.01$ and $A^*= 0.2\pm 0.02$.

\subsection{The 3D case}
In the 3D problem,  to introduce the ``immobilised'' wall a somewhat modified version of the \text{ParaDiS } code \citep{arsenlis2007enabling} was applied.
The \text{ParaDiS } approximates curved dislocations with a series of connecting short straight segments. During the evolution of the system, the connecting points of the segments are moved. The stress field at a segment is the sum of the ``mean field'' stress,  calculated by a fast multipole method (FMM) and the direct contribution of the nearby segments. The force acting on the segments is obtained according to the Peach-Koehler force, from which the velocity of the segments is given with an appropriate mobility functions. The dislocation segments are moved with a trapezoidal integrator using adaptive time-steps. Moreover, a topological algorithm handles intersecting dislocations and too long or too short segments.

In our simulations, single slip was considered. The cubic simulation  box was oriented so that the Burgers vector had only $x$ component ($\vec{b}=(b,0,0)$). The primary slip plane was rotated  by $20^{\circ}$ around the $x$ axis allowing the initial planar loops to spread into other parallel planes.
The box size was $14.377 \mu m$. The simulation parameters are summarised in Tab. \ref{tab:params}. The immobilised region with width $1.438 \mu m$ was introduced with a face parallel to the face $yz$ of the simulation box. The simulations were performed with periodic boundary conditions.

The external stress was applied only after all randomly generated systems were relaxed at zero external stress. In order to ensure flowing conditions, the external stresses applied were taken to be larger than the yield stress predicted by Taylor's formula, giving $\tau_{f}$ =5MPa.
\begin{table}[h]
    \centering
    \begin{tabular}{|c|c|}
        \hline
        $\mu$ & $64.88 GPa$ \\
        \hline
        $\nu$ & $0.3$\\
         \hline
         b & 0.287 $nm$ \\
          \hline
        loop size  & $19.409 \mu m$ \\
         \hline
        Number of loops & 10 \\
         \hline
        $\tau_0$ & $ 10 \ and \ 15MPa$  \\
         \hline
    \end{tabular}
    \caption{Parameters used in the simulations. The material parameters not mentioned are the  defaults of \text{ParaDiS }.}
    \label{tab:params}
\end{table}
For the calculation of the space dependent total and GND densities a custom made  Python code was developed.

\begin{figure}[pos=htbp]
\centerline{\includegraphics[height=8cm]{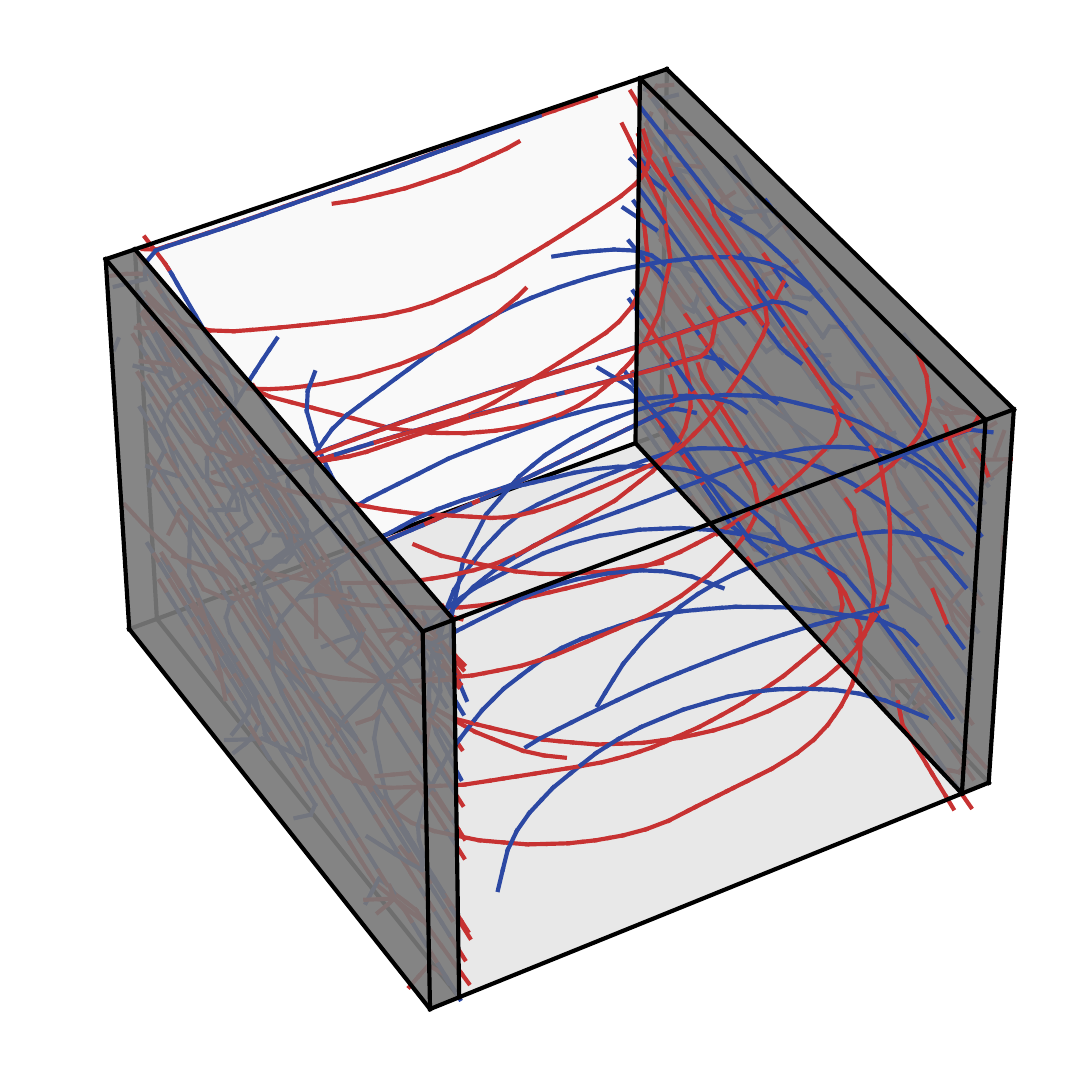}}
\caption{Typical 3D dislocation configuration with an impenetrable wall (gray area) in the simulation box at $\tau$=10MPa.}
\label{fig:3Dconf}
\end{figure}

\begin{figure}[pos=htbp]
 \includegraphics[height=8cm]{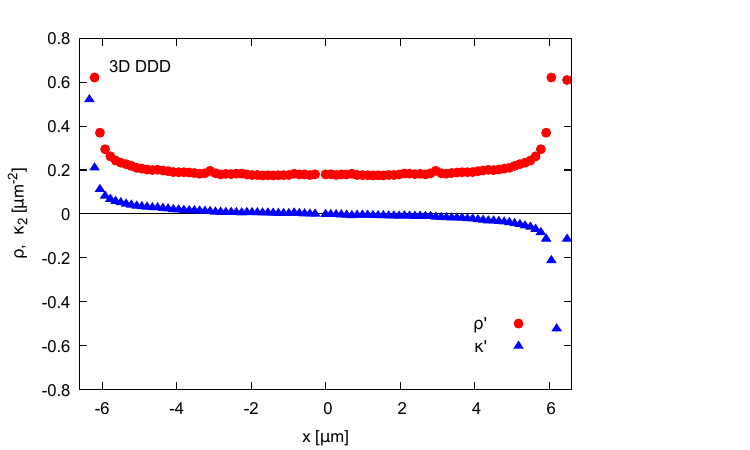}
 \caption{The statistically stored and the GND densities of dislocation as a function of the distance from the middle of the active zone. $\tau$=10MPa.} \label{fig:3D_rho}
\end{figure}

\begin{figure}[pos=htbp]
 \includegraphics[height=8cm]{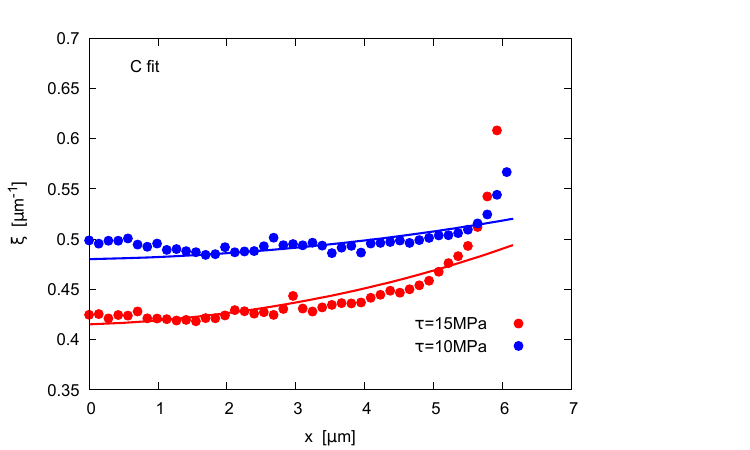}
 \caption{3D simulation results.  The $\xi'$ versus $x'$ plots at two different external loads. The fitting was performed for the data points corresponding to  $x<5.5 \mu m$. } \label{fig:3Dfit}
\end{figure}

\begin{figure}[pos=htbp]
 \includegraphics[height=8cm]{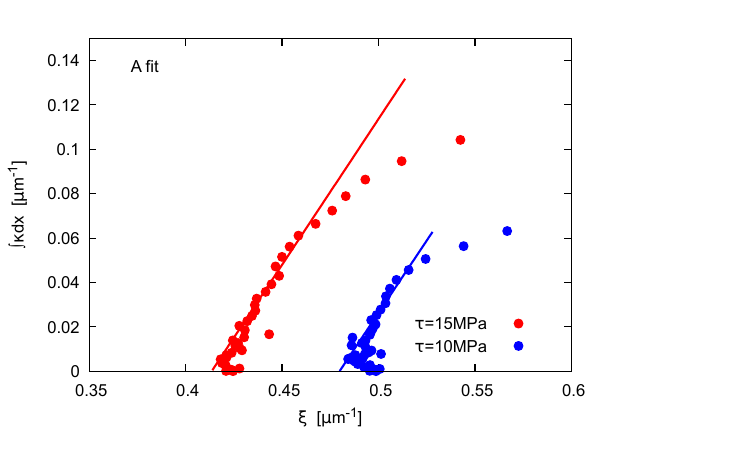}
 \caption{3D simulation results.  The $\int \kappa_2 dx$ versus $\xi$ plots at two different external loads. The fitting was performed for the data points corresponding to  $\int \kappa_2 dx<0.05 \mu m^{-1}$.} \label{fig:3Dkappafit}
\end{figure}

A typical dislocation configuration at $\tau_0=$10MPa is shown in Fig. \ref{fig:3Dconf}.
The functions $\rho(x)$ and $\kappa_2(x)$ obtained by averaging 100 simulations that started with different initial dislocation configurations are shown in Fig. \ref{fig:3D_rho}.
The curves $\xi$ versus $x$ with the fitted theoretical functions at two different external stress levels  are plotted in Fig. \ref{fig:3Dfit}. 
The curves $\int \kappa_2 dx$ versus $\xi$ for the two applied load levels are plotted in Fig. \ref{fig:3Dkappafit}.

It should be noted that, in contrast to the 2D case, the fitted curves do not follow the measured data points in a wider region next to the immobilised wall. This can be attributed to the fact that due to the high gradients of the dislocation density in this region, modelling the situation may require the introduction of higher order terms in the plastic potential \citep{groma2015scale} and the second term on the right hand side of Eq. (\ref{eq:tau_2}) may not be negligible.
Moreover, since the dislocation density is much larger next to the wall than in the middle region, the time needed to reach the steady state configuration is very much enhanced near the wall. Thus, it cannot be easily reached within a reasonable computation time. This issue requires further investigation.  So, to obtain reasonable fits  data points corresponding to a region near the wall were excluded.

The actual values of the parameters obtained are: $\alpha= 0.3$, $C= 0.17\pm 0.02$, and  $A^*= 0.4\pm 0.1$.  From Eqs. (\ref{eq:C}) one obtains $A^*D_{22}=13.7\pm 1.4$.

\section{Conclusions}
The properties of a generalised version of continuum theory of curved dislocations based on a scalar functional of the fields - statistically stored dislocation density, geometrically necessary dislocation density, and curvature- are discussed. The theory contains parameters that can be determined by comparing the DDD simulation results with the numerical solution of the continuum theory.

In the investigations presented, a narrow impenetrable wall is introduced into the simulation box. After applying an external load, an inhomogeneous dislocation distribution develops near the wall, with spatial variation depending on three parameters of the continuum theory. By a straightforward analytical calculation, one can see that, for this simple geometry, the solution of the continuum theory can be obtained through a simple numerical integration. Therefore, obtaining a solution does not require any sophisticated finite element method.

It was shown that the spatial variation of both the dislocation and GND densities obtained from 2D and 3D DDD simulations can be well fitted by the numerical solution of the continuum theory. The fitting gives three parameters ($D_{22}$, $A^*$, and $\alpha$) of the continuum theory.

Certainly, the simple numerical method needed to solve the rather specific problem considered cannot be applied to more complicated boundary value problems. For that, efficient finite element methods have to be developed. However, the " wall " problem can serve as a benchmark problem for testing the finite element code.

\printcredits

\section*{Declaration of competing interest}
The authors declare that they have no known competing financial interests or personal relationships that could have appeared to influence the work reported in this paper.

\section*{Acknowledgement}
The financial support of the Hungary National Research, Development and Innovation Office (IG and PDI, Project No. NKFIH EXCELLENCE25 153976) is acknowledged.  PDI is also supported by the János Bolyai Scholarship of the Hungarian Academy of Sciences.

\bibliographystyle{cas-model2-names}


\end{document}